\pdfoutput=1
\let\accentvec\vec
\documentclass[structabstract]{aa} 

\let\vec\accentvec
\let\epsilon\varepsilon
\let\theta\vartheta

\usepackage{graphicx}
\usepackage{rotating}
\usepackage{enumerate}
\usepackage[english]{babel}
\usepackage[T1]{fontenc}
\usepackage{hyperref}
\usepackage{amsmath}
\usepackage{longtable}
\usepackage[utf8]{inputenc} 
\usepackage{txfonts}
\usepackage{natbib}
\bibpunct{(}{)}{;}{a}{}{,}
\newcommand{\hess}{\textit{H.E.S.S.}}
\newcommand{\fermilat}{\textit{Fermi-LAT}}
\newcommand{\fermi}{\textit{Fermi}}
\newcommand{\iact}{\textit{IACT}}
\newcommand{\swiftxrt}{\textit{Swift-XRT}}
\newcommand{\swiftuvot}{\textit{Swift-UVOT}}
\newcommand{\atom}{\textit{ATOM}}
\newcommand{\tenten}{\textit{1RXS~J101015.9~-~311909}}

\newcommand{\mrk}[1]{\textit{Mrk~#1}}
\newcommand{\pks}{\textit{PKS~2155-304}}

\begin{document}

\titlerunning{Constraining the parameter space of the one-zone SSC model for GeV-TeV detected BL~Lac~objects}
%\titlerunning{Constraining the parameters of the SSC model for GeV-TeV BL Lacs}
   \title{Constraining the parameter space\\ of the one-zone synchrotron-self-Compton model\\ for GeV-TeV detected BL~Lac~objects}

   \author{M. Cerruti
          \inst{1,2}
          \and
          C. Boisson\inst{1}
                    \and
          A. Zech\inst{1}
          }

   \institute{LUTH, Observatoire de Paris, CNRS, Universit\'{e} Paris Diderot; 5 Place
Jules Janssen, 92190 Meudon, France    
          \and
          Harvard-Smithsonian Center for Astrophysics; 60 Garden Street, Cambridge MA, 02138, US  \\ 
          email: matteo.cerruti@cfa.harvard.edu}   
             
 \date{Received: December 20, 2012;  Accepted: }
 
   \abstract
  { The one-zone synchrotron-self-Compton (SSC) model aims to describe the spectral energy distribution (SED) of BL Lac objects via synchrotron emission by a non-thermal population of electrons and positrons  in a single homogeneous emission region, partially upscattered to $\gamma$-rays by the particles themselves.}
   {The model is usually considered as degenerate, given that the number of free parameters is higher than the number of observables. It is thus common to model the SED by choosing a single set of values for the SSC-model parameters that provide a good description of the data, without studying the entire parameter space. We present here a new numerical algorithm which permits us to find the complete set of solutions, using the information coming from the detection in the GeV and TeV energy bands.}
   {The algorithm is composed of three separate steps: we first prepare a grid of simulated SEDs and extract from each SED the values of the observables; we then parametrize each observable as 
   a function of the SSC parameters; we finally solve the system for a given set of observables. We iteratively solve the system to take into account uncertainties in the values of the observables, producing a family of solutions.}
   {We present a first application of our algorithm to the typical high-frequency-peaked BL Lac object \tenten, provide constraints on the SSC parameters, and discuss the result in terms of our understanding of the blazar emitting region.}
   {}

   \maketitle
   \section{Introduction}

  Blazars are a class of active galactic nuclei (AGN) characterised by extreme variability, a high degree of polarization and a strong non-thermal continuum observed in the optical/UV spectrum \citep{Stein76, Moore81}. According to the unified AGN model, they are considered as radio-loud AGN with the relativistic jet pointing in the direction of the observer \citep{Urry95}. Their spectral energy distribution (SED) is thus dominated by the non-thermal emission from the jet, enhanced by relativistic effects. The blazar class is divided into the two subclasses of flat-spectrum-radio-quasars (FSRQs) and BL Lacertae (BL Lac) objects according to the strength of the non-thermal continuum emission relative to the thermal emission from the accretion disc that is partially reprocessed in the broad-line-region (BLR). A blazar is then classified as a BL Lac object if the optical/UV spectrum is dominated by the continuum emission, while it is classified as an FSRQ if emission lines and/or the big blue bump are observed \citep[the standard threshold between the two subclasses being an equivalent width of the emission lines equal to $5 \ensuremath{\mathring{\text{A}}}$, ][]{Angel80}.\\
   The blazar SED is composed of two components, the first one peaking between infra-red and X-rays, the second one peaking in $\gamma$-rays \citep[see e.g.][]{FermiSED}. 
    The position of the peak frequency of the first component is used to differentiate among subclasses of BL Lac objects: we can then identify a BL Lac object as high-frequency-peaked BL Lac (HBL) if the peak frequency is located in UV/X-rays, or as low-frequency-peaked BL Lac (LBL) if the peak frequency is located in infrared/visible light \citep[see][]{Padovani95}. FSRQs always show a low-frequency peak. The position of the first peak frequency seems to be inversely correlated with the luminosity \citep[the so-called \textit{blazar sequence},][]{Fossati98}, even though there is not a general consensus on this point \citep[see e.g.][]{Nieppola08, Padovani12}.\\
     
    The origin of the first bump is generally ascribed to synchrotron emission from a non-thermal population of electrons and positrons in the emitting region, while the second bump is often ascribed to inverse Compton processes of the same leptons with their own synchrotron emission \citep[synchrotron-self-Compton model, SSC,][]{Konigl81} or with an external photon field \citep[external-inverse-Compton, EIC,][]{Dermer92, Dermer93, Sikora94} such as the BLR, the accretion disc or the external torus. While the SSC model successfully describes the SED of HBLs, an external Compton component is required for LBLs and FSRQs \citep[see e.g.][]{Ghisellini11}.\\ 
    It is worth recalling that other kinds of models exist, in particular hadronic models in which the emitting region contains also relativistic protons, which are responsible (together with the secondary particles coming from p-$\gamma$ interactions) for the high energy component of the SED \cite[see e.g.][]{Mannheim93, Mucke01}.\\
    
   The one-zone SSC modelling of HBLs is usually considered as degenerate: the number of free parameters is higher than the number of observational constraints, and a best-fit solution cannot be given. The standard approach is thus to find one set of parameters that successfully describes the SED, without evaluating the uncertainties on these parameters. In this paper we present a new algorithm to constrain the parameter space of the model, using the information from the GeV and TeV $\gamma$-ray emission, as observed by \fermilat\ \citep{Atwood09} and ground-based imaging atmospheric Cherenkov telescopes (\iact s), respectively.\\ 
      
   Before describing our algorithm, we recall here the basis of the one-zone SSC model by \citet{Katarzynski01}, on which this study is based. The algorithm can easily be adapted to similar models. The emitting region is considered to be a spherical blob of plasma (with radius $R$), moving in the relativistic jet with a Doppler factor $\delta$. The source is supposed to be filled with a tangled, homogeneous magnetic field $B$ and a non-thermal population of leptons (e$^\pm$). The particle energy distribution is described as a broken power-law function \citep[a spectral break is expected in presence of synchrotron radiation, see e.g.][]{Susumu}, defined by the two indexes $\alpha_1$ and $\alpha_2$, the three Lorentz factors $\gamma_{min}$, $\gamma_{break}$ and $\gamma_{Max}$ and the normalization factor $K$. The free parameters of the model are then nine: three for the emitting region ($R$, $\delta$ and $B$) and six for the particle population ($\gamma_{min;break;Max}$, $\alpha_{1;2}$ and $K$). In general there are less than nine observables to constrain the parameters of the model, which remains thus degenerate. The basis of the constraints on the SSC model are described in the work by\defcitealias{Tavecchio98}{T98}\citet{Tavecchio98} (hereafter \citetalias{Tavecchio98}). The idea is to determine a set of equations linking parameters and observables, and to solve it analytically. In their work, \citetalias{Tavecchio98} used six observables: the frequency and luminosity of the synchrotron peak, the frequency and luminosity of the Compton peak, and the measured X-ray and $\gamma$-ray spectral slopes ($\Gamma_{X,\gamma}$, linked directly to $\alpha_1$ and $\alpha_2$ through $\alpha_{1,2}=2\Gamma_{\gamma,X}-1$). The number of free parameters was reduced from nine to seven by considering a very low (very high) value of $\gamma_{min}$ ($\gamma_{Max}$), and additional constraints were added from information about the variability time-scale and from the requirement that the break Lorentz factor be consistent with the one expected from synchrotron cooling.\\
   
   In this paper, we present a new algorithm to find the complete set of solutions for the interpretation of SEDs of HBLs with a one-zone SSC model. Our method is based on the approach by \citetalias{Tavecchio98}, but is fully numerical. In the next Section we provide the details of our algorithm, showing in Section \ref{sec3} a first application to the HBL \tenten.

  \section{Description of the algorithm}
  \label{sec2}
  
  \begin{figure*}[t!]
	 \centering
   \includegraphics[width=18cm, angle=0]{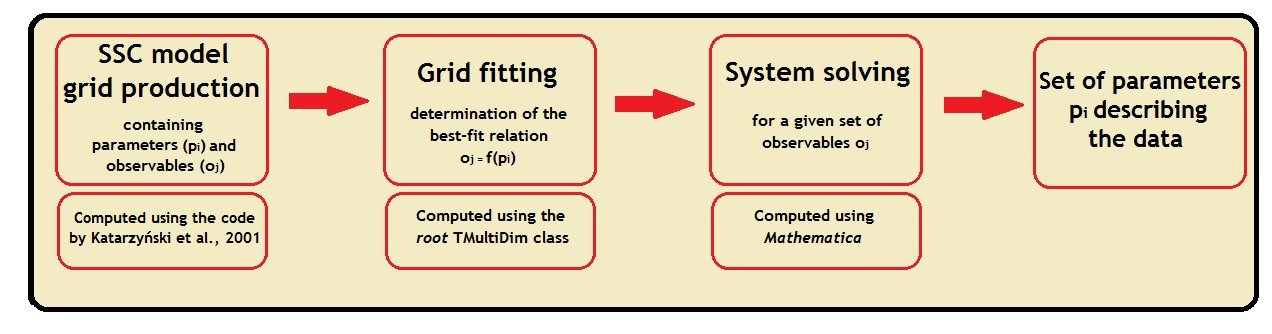}
     \caption[Flow diagram for the new numerical algorithm for constraining the SSC model parameters.]{Flow diagram for the new numerical algorithm for constraining the SSC model parameters.}
     \label{algochart}
   \end{figure*}

  Our algorithm can be seen as a numerical extension of the work done by \citetalias{Tavecchio98}. The basic idea is to define a set of equations linking SSC parameters and observables that, in our case, are obtained numerically. The algorithm is composed of three separate steps (summarized in Fig. \ref{algochart}):
  \begin{itemize}
  \item we simulate SEDs for a grid of parameter values (spanning the region of interest of the SSC parameters $p_j$), and for each SED we compute the value of the corresponding observables ($O_i$);\\
  \item we then parametrize each observable $O_i$ as a function of the SSC parameters: 
  \begin{equation}
  \label{equation1}
\log{O_i} = \sum_{j} \alpha_j\ \log{p_j}
\end{equation} 
where $\alpha_j$ is the result of a fit;\\
\item we finally solve the set of equations for a given set of observables $O_i$, to obtain the solution of our model $p_j$. If the observables $O_i$ include an intrinsic uncertainty ($\sigma_i$), we iteratively solve the system for $O_i\in[O_i-\sigma_i, O_i+\sigma_i]$, thus producing a set of possible solutions.\\
\end{itemize} 

We reduce the number of free parameters from nine to seven by fixing a reasonably low and high value for $\gamma_{min}$ (fixed to $100$) and $\gamma_{Max}$ (fixed to $5\cdot10^6$), respectively.\footnote{These values are consistent with the ones adopted by \citetalias{Tavecchio98}. For the particular case of \tenten\ presented in the next Section, values of $\gamma_{min}$ lower than 100 would overestimate the radio measurements.} The main difference with respect to the work by \citetalias{Tavecchio98} lies in the choice of the observables. The synchrotron peak is relatively well constrained. Even when it is located in the observational gap between the ultra-violet and X-ray bands, its position and luminosity can be reasonably well estimated by extrapolating the low and high energy data. The position of the inverse Compton peak is however much more uncertain.
 We thus decided, instead of the latter, to rather use the actually observed GeV and TeV spectral slopes, and their flux normalizations, defined in the following as $\Gamma_{GeV}$, $\Gamma_{TeV}$, $\nu F_{\nu;GeV}$ and $\nu F_{\nu;TeV}$. The number of observables we use is thus seven: together with the four $\gamma$-ray observables, we consider the frequency and intensity of the synchrotron peak ($\nu_s$ and $\nu F_{\nu;s}$), and the measured X-ray spectral slope ($\Gamma_X$). We thus have seven free parameters and seven independent observables, and the problem can be solved.\\

 It is important to underline that the four $\gamma$-ray observables are not degenerate: the shape of the inverse Compton component is \textit{not} symmetrical with respect to its peak, given the fact that the TeV part is affected by the transition to the Klein-Nishina regime, the internal absorption by $\gamma$-$\gamma$ pair production \citep[which depends on the synchrotron emission, see e.g.][]{Aha08} and the absorption on the extra-galactic background light \citep[EBL,][]{Salamon98}. It is clear that such a study is carried out more easily with a numerical approach (a purely theoretical derivation of the expression of the TeV spectral slope assuming all the effects described above would be non-trivial). Another improvement with respect to the work by \citetalias{Tavecchio98} is that we do not use the simple approximation that the GeV slope is uniquely related to the electron spectral index below the break (when $\gamma_{min}$ is reasonably low): this approximation is true only if the inverse Compton component accounts for the GeV spectrum  well before the peak; when the Compton peak is located at lower energies, the spectrum measured with \fermilat\ is significantly softer.\\

In the SSC code used for this study \citep{Katarzynski01}, a modification has been introduced concerning the definition of the normalization factor $K$ of the electron distribution, which is originally defined as the number density at $\gamma=1$. With this definition, the value of $K$ depends strongly on the value of $\alpha_1$: in particular, assuming that the model correctly fits the synchrotron peak, but we want to modify $\alpha_1$, this imposes a correction of the value of $K$, as well. For a simulation of a multitude of SEDs, this definition is not appropriate, imposing a huge range of values for $K$. We thus redefined $K$ as the number density at $\gamma=\gamma_{break}$ :
\begin{equation}
\label{equation2}
K'=K \gamma_{break}^{-\alpha_1}
\end{equation} 
In this way, if the model correctly describes the synchrotron peak, one can modify the value of $\alpha_1$ without affecting the other parameters, and, more importantly, the range of values of $K'$ to study becomes narrower. \\
As another small modification, the EBL absorption is computed using the model by \citet{Franceschini}.\\

In order to minimize the computing time, the code has been parallelized using \textit{OpenMP}\footnote{\url{www.openmp.org}}. The parallelization is not at the level of the computation of the synchrotron and inverse Compton components, but at the higher level of the sampling of the parameter space. The distribution of the input parameters per thread is done dynamically, and the communication between threads takes place only at the end of each computation, when the results are merged.\\ 

Iterating on the values of $\delta$, $B$, $K'$, $R$, $\gamma_{break}$ and $\alpha_1$, SEDs have been simulated for a grid of parameter values (\textit{''grid production''} in Fig. \ref{algochart}). The value of $\alpha_2$ was fixed, constrained by the measured X-ray spectral slope $\Gamma_X$, and equal to $2\Gamma_X - 1$. For each modelled SED, we first identify the position of the synchrotron peak and its intensity. To compute the expected fluxes and spectral slopes detected in the GeV and TeV ranges, we simply fit the modelled SED in the range of the GeV and TeV detection with a power-law function, using as minimum and maximum energies the values adopted in the spectral fitting of the \fermilat\ and \iact\ detection.   This defines the two indexes $\Gamma_{GeV;TeV}$, and the two fluxes (of the fitted power-law, not of the model) $\nu F_{\nu;GeV;TeV}$ at the respective decorrelation energies (which are measured quantities).\\ 
For the application that will be presented in the next Section, we sampled $5$ different values for each free parameter, resulting in the computation of $5^6=15625$ SEDs. As an indication, the computing time on our $16$-core machine is roughly one hour. The result of this stage is a grid containing, for each set of parameters, the corresponding set of observables.\\

This grid has then been fitted using as dependent variables the six observables, and as independent variables the six free parameters (\textit{''grid fitting''} in Fig. \ref{algochart}). While for the synchrotron peak frequency and flux the relation between observables and parameters is simple (and consistent with the analytical expression), for the fluxes and slopes measured in the GeV and TeV energy bands, it becomes more complicated. In particular, the simple relation given in Equation \ref{equation1} does not apply any more, and we need to consider a more complex relation of the form
\begin{equation}
\label{equation3}
 \log{O_i} = \sum_{k1...k6}{\alpha_{k1...k6}\prod_{j}{\log{p_j}^{k1...k6}}}
\end{equation} 
 More explicitly, we have to perform a fit considering all the possible polynomials of our parameters.\\
The problem is then to find a parametrization for a dependent variable, function of six parameters, without knowing it \textit{a priori} (nonparametric regression). This problem has been solved using the \textit{root} software\footnote{\url{http://root.cern.ch/}}, namely the \textit{TMultiDim} class. This class provides, for each term of the fit function, the fitted coefficient and the weight of the term in the overall fit, providing a listing of the different terms in order of relative importance.\\
The computing time for this step is a few seconds for the synchrotron observables and for $\nu F_{\nu;GeV;TeV}$, while for the GeV and TeV slopes it can be significantly longer, taking up to an hour each if we want to study all the possible polynomial functions (in order to be sure not to miss a high-order polynomial which might play an important role in the fit). The choice of the last polynomial considered in the fit is a free parameter of the algorithm: it is chosen by listing the fit terms according to their contribution to the $\chi^2$, and defining a threshold above which the contribution of further terms becomes negligible. The goodness of the fit has thus to be verified, as shown in Fig. \ref{1RXSJ1010fits} for the application presented in the next Section.\\

Once we have obtained the six equations relating the observables to the free parameters, the final stage of the algorithm is to solve this system of equations (\textit{''system solving''} in Fig. \ref{algochart}). This task is carried out with the \textit{Mathematica} software\footnote{\url{http://www.wolfram.com/mathematica/}}, using the numerical \textit{FindInstance} command, and the solutions have been reduced to the real domain. The computing time for this step depends strongly on the form of the equations in our system. In order to reduce it, we have fixed one of the parameters ($\alpha_1$ for the application presented in the next Section), and searched the solutions of the system for different given values of this frozen parameter. For each given value, we solve a system of five equations plus two inequalities, corresponding to the minimum and the maximum of $\Gamma_{GeV}$\footnote{There is no constraint on the choice of the frozen parameter ($\alpha_1$ in this case), nor on the choice of the equation which is turned into two inequalities ($\Gamma_{GeV}$ in this case) to maintain the same number of equations and variables.}. An additional inequality is added to the system, relating the size and the Doppler factor to the variability time-scale $\tau_{var}$:
\begin{equation}
\label{vareq}
\tau_{var}\geq \frac{1+z}{c}\  \frac{R}{\delta}
\end{equation}
where $c$ is the speed-of-light and $z$ is the redshift of the source.\\

 To take into account the uncertainty on the five remaining observables, we iterate the solution of the system spanning the range $O_i\in[O_i-\sigma_i, O_i+\sigma_i]$, to produce a set of solutions. We sampled three different values of the frozen parameter, and we span the five observables 12 times each, leading to $3\cdot12^5=746496$ systems studied. The \textit{Mathematica} code has been parallelised as well, and run on a computer grid, using up to 200 cores simultaneously. For the expressions used for the application presented in the next Section, the computing time is roughly nine hours.\\

The three steps of the analysis (grid production, grid fitting and system solving, resumed in the plot shown in Figure \ref{algochart}) can then take several hours, and this imposes a compromise between the resolution adopted for the grid (which affects the computing time of both step one and two), the number of terms used in the fit (which affects the computing time of both step two and three), and the number of iterations done in the system solution (which affects the computing time of step three).\\  
 
 The sets of parameters found with our algorithm are then used to produce contour plots, showing the confidence regions as a function of two different parameters, as in Fig. \ref{1RXSJ1010contourplot}. As a final step, we compute all the SSC models corresponding to the found solutions, and we evaluate for each of them the reduced chi-square ($\tilde{\chi^2}$) with respect to the observational data. This check allows us to estimate the true best-fit solution (the one with the minimum $\tilde{\chi^2}$ value), and to verify that the contour plots correspond indeed to $1$-$\sigma$ contours (i.e. that all the solutions are characterised by $\Delta \tilde{\chi^2} \leq 2.3$ with respect to the best-fit value).

\begin{table*}[t!]
\begin{center}
\caption[Summary of the constrained parameters for the SSC modelling of \tenten.]{Summary of the constrained parameters for the SSC modelling of \tenten. The solutions have been computed for three different values of $\alpha_1$, which have been frozen in the solution of the system. For the other five free parameters, we provide the minimum and the maximum values obtained, corresponding to the $1$-$\sigma$ confidence interval. The solutions with a Doppler factor higher than $100$ have been excluded. Note that the model parameters are correlated, as shown in the contour plots (Fig. \ref{1RXSJ1010contourplot}).\vspace{0.5cm}}
\label{1010fittable}
\begin{tabular}{|c||c|c|c|}
\hline
\hline
$\alpha_1$ & 1.6 & 1.8 & 2.0 \\[1pt]\hline  \\[-1.0em]
$\gamma_{break}$ & $(3.26 - 9.62)\cdot10^4$ & $(3.43-11.21)\cdot10^4$ & $(3.49-13.14)\cdot10^4$ \\[1pt] \hline \\[-1.0em]
$B$ [G] & $(1.00-3.86)\cdot10^{-2}$ & $(0.77-3.86)\cdot10^{-2}$ & $(0.51-4.08)\cdot10^{-2}$ \\[1pt]\hline  \\[-1.0em]
$K'$ [cm$^{\textrm{-3}}$] & $(0.061-14.70)\cdot10^{-7}$ & $(0.023-9.92)\cdot10^{-7}$ & $(0.0093-7.30)\cdot10^{-7}$\\[1pt]\hline  \\[-1.0em]
$\delta$ & $30.89-99.64$ & $30.76-99.91$ & $32.08-99.55$ \\\hline  \\[-1.0em]
$R$ [cm] & $(0.43-8.28)\cdot10^{16}$ & $(0.46-9.66)\cdot10^{16}$ & $(0.49-11.57)\cdot10^{16}$ \\[1pt]\hline\hline
\end{tabular}
\end{center}
\end{table*}

 \begin{figure*}[htbp!]
	 \centering
   \includegraphics[width=19cm, angle=0]{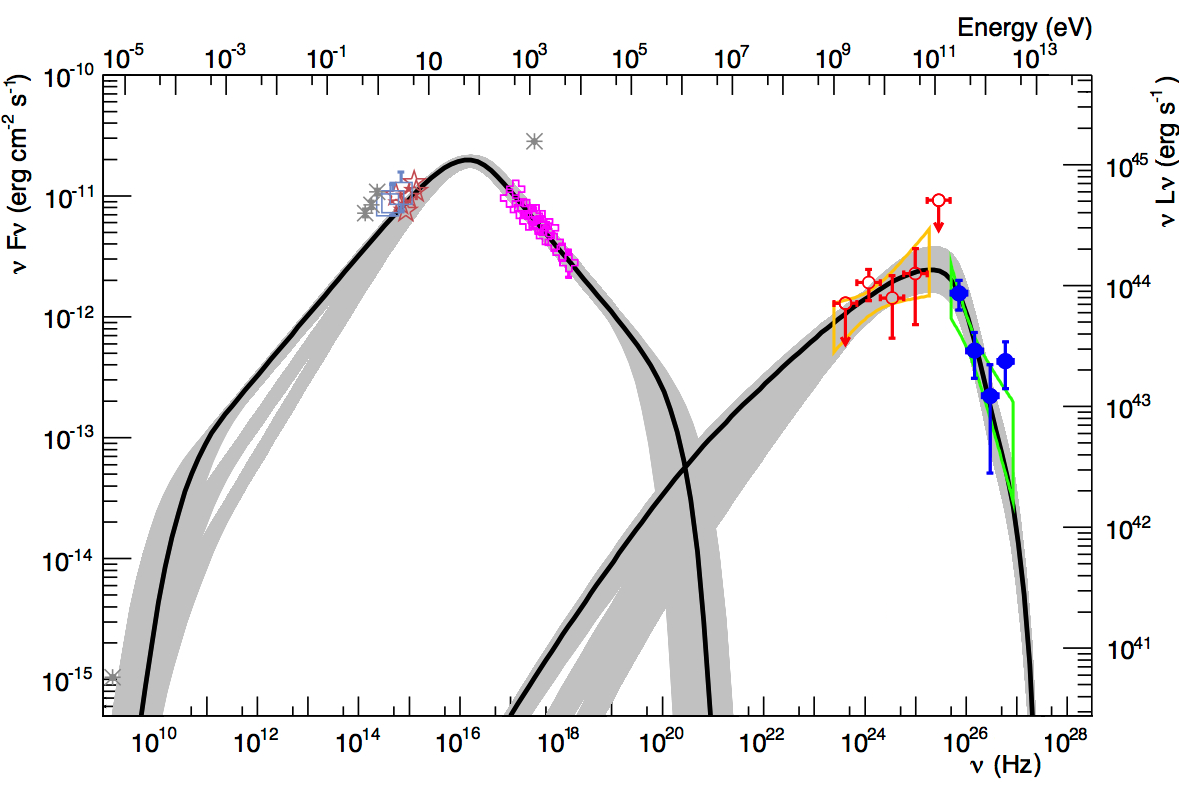}
     \caption[SED of \tenten.]{SED of \tenten\ \citep[][; the \hess\ spectrum is represented by the green bow-tie and the blue points; the \fermilat\ spectrum is represented by the orange bow-tie and the red empty circles; \swiftxrt\ data are shown by the pink crosses; \swiftuvot\ data are shown by the red stars; \atom\ data are shown by the blue open boxes; archival data from the NED are shown in grey]{tentenpaper}. In grey are plotted all the SSC models which describe the SED, as found with our algorithm, while the solid black curve represents the best-fit solution, with $\tilde{\chi^2}=1.06$. It is characterised by an extreme value of  $\delta$=$96.83$, $B$=$0.015$ G, $R$=$1.3\cdot10^{16}$ cm, $\alpha_1$=$2.0$, $K'$=$8.94\cdot10^{-8}$ cm$^{-3}$ and $\gamma_{br}$=$5.31\cdot10^{4}$.  The three different families of solutions, which can be distinguished in the range between $10^{11}$ and $10^{14}$ Hz, correspond to $\alpha_1$=$1.6$, $1.8$ and $2.0$, as discussed in Section \ref{sec3}. The infrared and visible data can be reproduced by taking into account the host-galaxy contribution.}
     \label{1RXSJ1010AllSol}
   \end{figure*}

  \section{Application: \tenten}
  \label{sec3}
    
  In this Section we present a first application of our numerical algorithm to the typical HBL \tenten. The TeV emission from this source, located at a redshift of $0.143$ \citep{Piranomonte07}, has been detected with \hess\ \citep{tentenpaper}. The multi-wavelength study presented in the \hess\ discovery paper shows that there are two major systematic uncertainties in the SED of this object. The first one comes from a possible absorption effect in X-rays, affecting the evaluation of the position of the synchrotron peak; the second one comes from a problem in the determination of the \fermi\ spectrum at low energies due to a possible contamination by diffuse $\gamma$-ray emission from the Galactic foreground. In the following we consider that the synchrotron peak is located at an energy between the available UV and X-ray data \citep[case B in ][]{tentenpaper}, and the \fermi\ spectrum is evaluated using a $1$ GeV low-energy threshold. This does not represent a limitation of the algorithm, which can be successfully applied to the other cases discussed by \citet{tentenpaper} as well.\\  
  
     The value of $\alpha_2$ is uniquely constrained by the X-ray spectrum measured with \swiftxrt\ ($\Gamma_X=2.5 \pm 0.1$) and it has been fixed at $4.0$. The grid of simulated SEDs has been produced sampling the other six free parameters in the ranges: $\delta\in[20,100]$, $B\in[0.005, 0.05]$ G, $\alpha_1\in[1.5,2.2]$, $\gamma_{break}\in[3\cdot 10^4,1.3\cdot10^5]$, $K'\in[10^{-9}, 2\cdot10^{-6}]$ cm$^{-3}$, and $R\in[4\cdot10^{15}, 10^{17}]$ cm. As laid out in the previous Section, the grid has been produced sampling five different values for each parameter, logarithmically spaced (apart from $\alpha_1$ which has been sampled linearly) between the minimum and the maximum values. It is important to underline that, solutions found \textit{outside} of the sampled grid, are not considered, given the fact that the system of equations is non-linear, and the fitted expression of the $O_i$ is appropriate only for the sampled parameter space. This could be seen as a limit of our approach, in the sense that we preselect a fraction of the parameter space. However, the analytical constraints defined by \citetalias{Tavecchio98} allow us to already exclude regions of the parameter space that are not accessible in the one-zone SSC model. In particular we decided not to study values of $\delta$ higher than $100$, which would require extreme values of the bulk Lorentz factor of the emitting region.\\

The result of the parametrization of the six observables is presented in the Appendix, in Tables \ref{cons1010sync} (for $\nu_s$ and $\nu F_{\nu;s}$), \ref{cons1010fgamma} (for $\nu F_{\nu; GeV, TeV}$), \ref{cons1010gFermi} (for $\Gamma_{GeV}$) and \ref{cons1010gHESS} (for $\Gamma_{TeV}$). As can be seen, while the expressions for $\nu_s$ and $\nu f_{\nu;s}$ are simple, and consistent with what is expected from analytical considerations, the expressions for the \fermi\ and \hess\ observables are more complicated, and we are obliged to consider second-order terms in the fitting polynomial, with the extreme case of $\Gamma_{TeV}$ which is described satisfactorily only when using more than one hundred terms. In Fig. \ref{1RXSJ1010fits} we show the relation between the values of each observable, and its reconstructed values (after fit): in a perfect fit, the values should follow exactly the linear relation, represented by the thick blue line.\\

\begin{figure*}[htbp!]
	 \centering
   \includegraphics[width=19cm, angle=0]{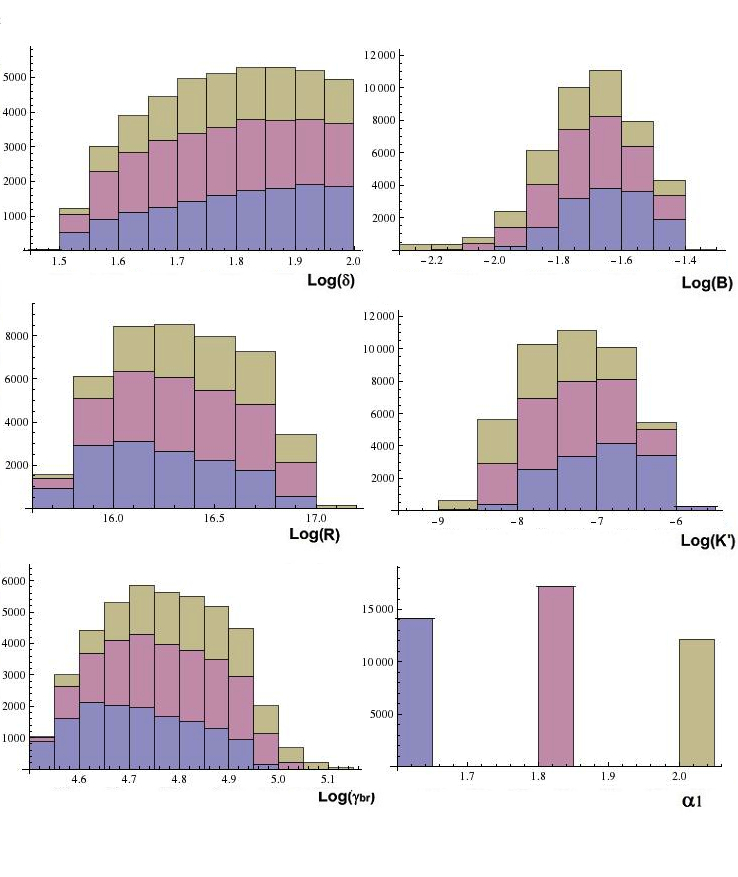}
     \caption[Solutions of the SSC modelling of \tenten\ obtained with our algorithm.]{Histograms showing the values of the SSC model parameters for the case of \tenten. From top to bottom, and left to right, we show the distribution of the solutions for $\delta$, $B$ (in G), $R$ (in cm), $K'$ (in cm$^{-3}$), $\gamma_{break}$, and $\alpha_1$. Please notice that the value of $\alpha_1$ has been frozen, and studied for three different cases ($1.6$, $1.8$ and $2.0$). The three colours represent the different solutions for the three values of $\alpha_1$ studied: violet-$1.6$, pink-$1.8$, and yellow-$2.0$.}
     \label{1RXSJ1010results}
   \end{figure*}

The system of equations was then solved for three different values of $\alpha_1$ (equal to $1.6$, $1.8$ and $2.0$) and for twelve different values of each observable (apart from $\Gamma_{GeV}$, which is defined through two inequalities), reaching from $O_i-\sigma_i$ to $O_i+\sigma_i$. The range of values for $\alpha_1$ corresponds to expectations from standard acceleration scenarios. The values adopted for the observables are (in logarithm): $\nu_s\in[16.15,16.25]$, $\nu f_{\nu;s}\in[-10.74,-10.68]$, $\nu f_{\nu;GeV}\in[-11.98,-11.78]$, $\nu f_{\nu;TeV}\in[-12.36,-12.10]$, $\Gamma_{GeV}\in[-1.94,-1.48]$ and $\Gamma_{TeV}\in[-3.55,-2.61]$. For the $\gamma$-ray observables, the considered error does include the systematic error on the observable (summed in quadrature to the statistic one). The system of equations includes an inequality for the variability time-scale ($\tau_{var} = 24$ hours) plus a limit on the value of the Doppler factor ($\delta<100$, which corresponds to the limit of the sampled grid).\\
The results of our algorithm, defining the range of SSC parameters which correctly fits the data, are shown in Table \ref{1010fittable} and the corresponding histograms are shown in Fig. \ref{1RXSJ1010results}.\\
In order to compare our solutions with the observational data, in Fig. \ref{1RXSJ1010AllSol} we show the SED of \tenten, together with all the modelled SEDs found with our algorithm. The $\tilde{\chi^2}$ of each modelled SED is estimated taking into account X-ray and $\gamma$-ray data only (i.e. excluding optical/UV and archival data), and the best-fit solution, shown in black in Fig. \ref{1RXSJ1010AllSol}, has $\tilde{\chi^2}=1.06$.\\
In Fig. \ref{1RXSJ1010contourplot} we show the contour plots corresponding to our solutions, for four different pairs of parameters: $B$-$\delta$, $R$-$\delta$, $\gamma_{br}$-$B^2 R$ and $u_e$-$u_B$ (the particle and magnetic field energy densities). The different sets of solutions that can be seen in the contour plots are due to the fact that the system of equations is solved for discrete values of the observables, in the range $\pm 1\sigma$. The only contour which is statistically significant is the most extended one, which represents the $1$-$\sigma$ confidence region.

Our algorithm, unlike the standard modelling approach, permits us to provide general, quantitative constraints on the physical parameters of the emitting region. These are discussed in the following for the case of \tenten.\\

\subsection{Doppler factor}

The first important result is that we can define a minimum Doppler factor that solves the system: $\delta_{min}\approx31$. In Fig. \ref{1RXSJ1010contourplot} (top-left plot) we show the contour plots of all our solutions in the $B$-$\delta$ plane, and we compare them to the analytical constraints computed following \citet{Tavecchio98}. It becomes clear that with our numerical approach we significantly narrow down the accessible parameter space.\\

The constraints on $\delta$ come mainly from the TeV slope and the variability time-scale: solutions with a lower Doppler factor exist, but would imply a variability time-scale higher than the one observed. In particular, comparing our result with the one presented by \citet{tentenpaper}, the fact that they provide a solution with $\delta=30$ is due to both a variability time-scale slightly higher than what has been assumed here ($\tau_{var}\approx25$ hours) and a rather soft TeV slope of the SSC model.\\ 

\subsection{Size of the emitting region}

\begin{figure*}[htbp!]
	 \centering
   \includegraphics[width=19cm, angle=0]{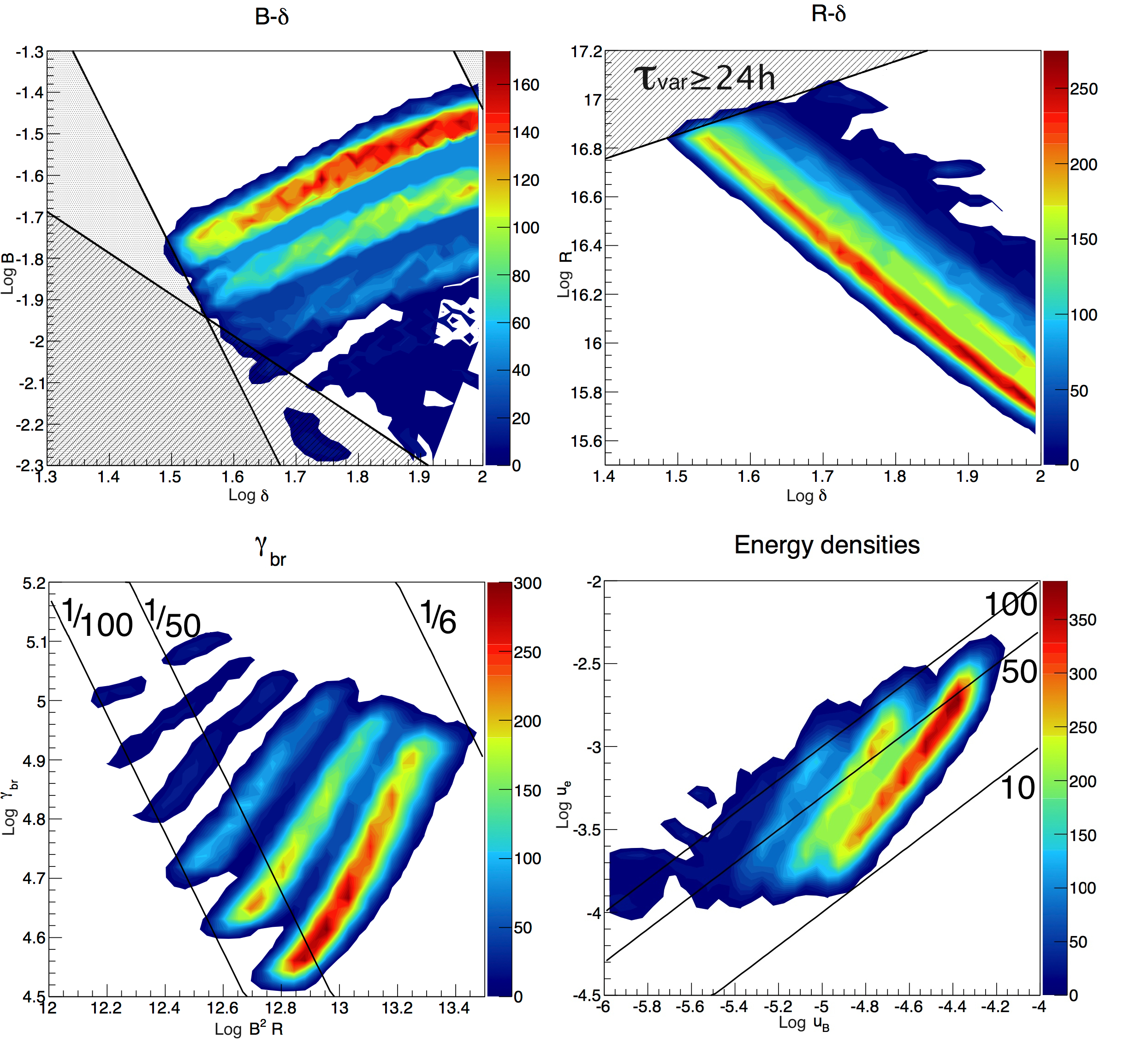}
     \caption{Contour plots of the solutions found for \tenten. The contours are expressed in arbitrary units, representing the number of solutions corresponding to a given pair of parameters. White represents zero. As discussed in Section \ref{sec2}, the most extended contour corresponds to a $1$-$\sigma$ contour with respect to the best-fit solution, as determined \textit{a posteriori}. \textit{Top left}: contours in the $B$-$\delta$ plane. The shadowed regions represent the exclusion-regions defined following \citetalias{Tavecchio98} (the region filled with diagonal lines is computed from Eq. 4, while the region filled with dots is computed from Eq. 11, considering an emitting region size equal to $(1+z)c\tau_{var} /\delta$ and $(1+z)c\tau_{var} / 30\delta$). We do not include the constraint on the consistency of $\gamma_{break}$ with synchrotron cooling). \textit{Top right}: contours in the $R$-$\delta$ plane. The filled region corresponds to a variability time-scale higher than $24$ hours. \textit{Bottom left}: contours in the $\gamma_{break}$-$B^2 R$ plane. The black lines represent the expected value of $\gamma_{break}$ in presence of pure synchrotron cooling, assuming a value of $\beta_{esc}$ equal to (from left to right) $1/100$, $1/50$ and $1/6$.   \textit{Bottom right}: contours in the $u_e$-$u_B$ plane. The black lines correspond to $u_e/u_B$ values equal to (from bottom to top) $10$, $50$, and $100$, respectively.}
     \label{1RXSJ1010contourplot}
   \end{figure*}

The size of the emitting region is limited by the variability time scale (Equation \ref{vareq}), following the usual considerations in SSC modelling.
The maximum allowed size is $\approx 10^{17}$ cm (at the edge of the values used to sample the grid). In Fig. \ref{1RXSJ1010contourplot} (top-right plot) we show the contour plots in the $R$-$\delta$ plane, as well as the limit corresponding to Equation \ref{vareq}. As discussed above, solutions with a larger size of the emitting region and a lower Doppler factor exist, but violate the variability constraint. For the case of \tenten, the variability constraint is relatively weak (no flaring behaviour observed), but for rapidly flaring sources this would permit to constrain the parameter space even tighter.\\

\subsection{Energy budget}

An open question in blazar physics concerns the energy budget of the emitting region. For each solution we compute the magnetic energy density ($u_B=B^2/8\pi$ in CGS units) and the particle energy density ($u_e=m_ec^2\int d\gamma\ \gamma N(\gamma)$). In Fig. \ref{1RXSJ1010contourplot} (bottom-right plot) all the solutions we found are shown in the $u_e$-$u_B$ plane, indicating that the system is significantly out of equipartition, with a $u_e/u_B$ ratio comprised between $10$ and $200$. The evaluation of $u_e$ depends on our assumptions on the low-energy part of the particle population (which is not constrained by the data), and in particular on the value of $\gamma_{min}$. A higher value of $\gamma_{min}$ would lower the value of $u_e$, reducing the equipartition-ratio. This effect has not be tested here, given the fact that high values of $\gamma_{min}$ affect the evaluation of the GeV slope, and would require a dedicated grid of SED model curves. However, for values of $\alpha_1$ harder than $2.0$, we do not expect that high values of $\gamma_{min}$ affect significantly this result, while it should be important when $\alpha_1 \geq 2.0$.\\

The fact that the SSC modelling of HBLs requires that the emitting region is out of equipartition is consistent with the results obtained by several authors (see e.g. \citet{Mrk421} for the case of \mrk{421}, \citet{Mrk501} for the case of \mrk{501} and \citet{pksmwl} for the case of \pks).

\subsection{Particle energy distribution}

 The study of the solutions for different values of $\alpha_1$, reveals that the slope of the electron distribution is not as well constrained by the GeV data as it is generally assumed \citepalias[and as in][]{Tavecchio98}. The solutions for the different $\alpha_1$ values are quite similar. An additional constraint on the values of the index of the particle population can be provided by optical/UV measurements. In the case of \tenten, however, the low-energy data cannot be used to define a unique $\alpha_1$ value, because of the uncertainty on the correction by absorption and by the host-galaxy contamination \citep[see ][and Fig. \ref{1RXSJ1010AllSol}]{tentenpaper}. For this source we can use the optical and UV data only as upper limits, excluding slopes softer than $2.2$. For other blazars, with a better estimate of the AGN emission in this part of the spectrum, a more precise value of $\alpha_1$ can be determined.\\

  The values of $\alpha_1$ that we have considered, point to one of the open questions of blazar physics: synchrotron cooling predicts a spectral break equal to one (i.e. $\alpha_1=3.0$, for $\alpha_2=4.0$, as constrained by the X-ray observations), which is clearly excluded by the data. This might indicate that the homogeneous one-zone model is too simple, and that more complex effects should be taken into account, such as non-linear inverse Compton cooling, energy-dependent acceleration and escape from the emitting region, non-homogeneity of both particles and magnetic field. This justifies the choice of considering the two particle slopes $\alpha_1$ and $\alpha_2$ as independent parameters.\\

 If the spectral break was a result of pure synchrotron cooling, the energy of the break in the particle distribution could be computed theoretically, and linked to the other free parameters. Assuming that particles are injected in the emitting region, are escaping from it at a speed $\beta_{esc}c$ \footnote{The particle escape term can be considered equivalently as an adiabatic term (i.e. associated to the expansion of the emitting region).}, and are losing their energy while emitting synchrotron radiation, the value of $\gamma_{break}$ can be expressed as \citepalias[see ][, Eq. 30]{Tavecchio98}:
\begin{equation}
\gamma_{break} = \beta_{esc}c \frac {5\cdot10^8}{B^2 R}
\end{equation} 
In Fig. \ref{1RXSJ1010contourplot} (bottom right plot) we show the values of $\gamma_{break}$ we found, as a function of $B^2 R$, comparing them to the theoretical expectations. As can be seen, our solutions are consistent with a pure synchrotron cooling only if the escape speed is lower than $c/6$. This value can be increased by considering additional cooling terms, as the energy loss by inverse Compton scattering.\\
This point, strengthens the fact that the observed spectral break is not consistent with pure synchrotron cooling, and that more complex effects have to be considered to explain the particle spectrum in the stationary state.\\

In this application we have fixed the values of $\gamma_{min}$ and $\gamma_{Max}$, which cannot be constrained in the particular case of \tenten. For other sources, with different observational constraints, it would be possible to fix other parameters (such as  $\alpha_{1}$, in presence of precise measurements at lower frequencies) and to study $\gamma_{min}$ or $\gamma_{Max}$.  In particular, this approach would be interesting for some other HBLs, for which a value of $\gamma_{min}$ higher than what we have assumed here is required in order to describe the SED with a SSC model \citep[see e.g.][]{Katarzynski06, Sarah}.

\section{Conclusions}

In this paper we have described a new algorithm to constrain the parameter space of the one-zone SSC model, which can be successfully applied to any HBL with simultaneous spectral measurements in the X-ray, GeV and TeV energy-range. The algorithm follows the idea developed by \citetalias{Tavecchio98} (i.e. the definition of a set of equations linking the free model parameters and the blazar observables) but it is fully numerical, introduces as new observables the properties of the GeV and the TeV detection, and allows the derivation of a new set of equations.\\
The algorithm cannot be applied to the subclasses of LBLs and FSRQs, given the fact that for these sources an external inverse Compton component is required, increasing the number of model free parameters.\\  

As a first test, we have applied our algorithm to the typical HBL \tenten\ and derived the set of solutions which correctly describes its SED. It is important to underline that the range of solutions found here is much narrower than the one obtained with the analytical approach, showing that the information coming from the GeV-TeV detection permits to better constrain the model. In this way, a true best-fit solution is provided.\\

Other numerical algorithms have been proposed in the literature to constrain the SSC model parameter space. Both \citet{Finke08} and \citet{Manku11ref} have proposed a $\chi^2$ minimization algorithm to fit the SEDs of BL Lac objects, showing as well that, thanks to the TeV detection, it is possible to obtain a best-fit solution. None of these two methods make use, however, of the recent \fermi\ data. On the other hand, their fits are strongly constrained by the low energy (optical and UV) photons, implicitly assuming that the $\gamma$-ray emitting region is responsible for the totality of this flux. Our method, based on the properties of the \fermi\ detection and the position of the synchrotron peak, allows us to relax this hypothesis, and to model more complex scenarios in which the low-energy flux is contaminated by emission from other, farther regions of the jet, or the host galaxy. It is worth to underline as well that the method described by \citet{Finke08} introduces the requirement that the energy budget of the emitting region is minimized, pre-selecting a set of solutions.\\

Several improvements to the algorithm we have presented here can be considered. First of all, the production of the grid of modelled SEDs, as was discussed here for the application to \tenten, limits the parameter space: a larger grid can be produced, extending the region of the parameter space under investigation.\\
To reduce the computing time we have also frozen the value of $\alpha_1$, solving the system for different, given values of this parameter: however the system can in principle be solved leaving all the parameters free to vary, studying $\alpha_1$ as a free parameter as well.\\

In most of the current publications, the standard approach in blazar physics is still to consider the one-zone SSC model as degenerate (i.e. the number of free parameters is higher than the number of constraints), and a fit of the SED is usually \textit{not} performed. It is thus common to explore only a few solutions to decide whether the data can be described in a satisfactory way, without studying the entire parameter space, nor evaluating the errors on the parameters. We have shown here that, at least for GeV-TeV HBLs, it is possible to explore the parameter space more systematically and to significantly improve the constraints on the parameters of the model. Our algorithm can be applied to the continuously increasing sample of HBLs detected in $\gamma$-rays, and will help us improve our comprehension of this extreme class of blazars.\\

  \section*{Acknowledgments}  

The work has been performed using the computation facilities of both the Paris Observatory and the Harvard-Smithsonian Center for Astrophysics. The authors wish to thank Helene Sol for fruitful discussions, as well as the anonymous referee for the comments and remarks which improved the present work.

 \bibliographystyle{aa}
  \bibliography{constraints_biblio}

   \begin{appendix}
  \section{Tables and figures containing the fit results}

\begin{figure*}[hbtp]
	 \centering
   \includegraphics[width=18.5cm, angle=0]{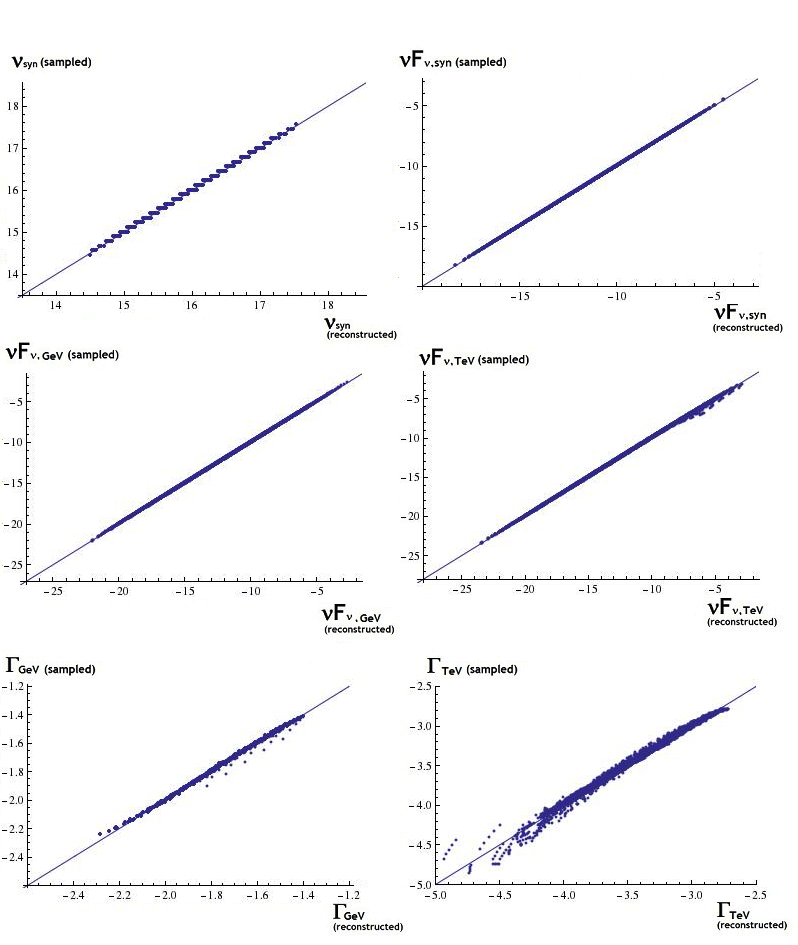}
     \caption[Comparison between the sampled values of the six observables considered in this study and their reconstructed values for the case of \tenten.]{Comparison between the sampled values of the six observables considered in this study and their reconstructed values, expressed as a function of the SSC-model parameters, for the case of \tenten. In a perfect fit, the points would follow a linear relation (tiny solid line). The six subplots are in the order (from top to bottom, from left to right): the synchrotron peak frequency ($\nu_s$); the synchrotron peak flux (expressed as $\nu F_{\nu;s}$); the \fermi\ flux (measured at the decorrelation energy and expressed as $\nu F_{\nu;GeV}$); the \hess\ flux (measured at the decorrelation energy and expressed as $\nu F_{\nu;TeV}$); the measured \fermi\ photon index ($\Gamma_{GeV}$) and the measured \hess\ photon index ($\Gamma_{TeV}$).}
     \label{1RXSJ1010fits}
   \end{figure*}

  We present here the results of the fitting of the grid for the particular case of \tenten. For each observable we provide the terms of the fit listed by increasing polynomial order of the parameters composing each term. The plot showing the goodness of each fit (comparison between the sampled and the reconstructed values) are given in Fig. \ref{1RXSJ1010fits}.

\newpage    
  \begin{table*}[h!]
\begin{center}
\caption[Coefficients of the fit performed to obtain an expression of $\nu_s$ (left) and $\nu F_{\nu;s}$ (right) for the case of \tenten.]{Coefficients of the fit performed to obtain an expression of $\nu_s$ (left) and $\nu F_{\nu;s}$ (right) for the case of \tenten. The corresponding plots are the top plots of Fig. \ref{1RXSJ1010fits}. Each line corresponds to a term of the fit: the first column gives the total order of the polynomial, the next six columns give, for each model free parameter, the degree considered, and the last column gives the associated coefficient. The terms are listed with respect to the parameters composing each term, and then by increasing degree, \textit{not} by relative weight in the fit. In this case, the value of $\alpha_2$ has been fixed to $4.0$.}
\label{cons1010sync}\begin{tabular} {|c|c|c|c|c|c|c||c|||c|c|c|c|c|c|c||c|}
\hline
\multicolumn{8}{|c}{{$\nu_s$}}  & \multicolumn{8}{c|}{{$\nu F_{\nu;s}$}}\\
\hline
\hline
Order & $\gamma_{br}$ & $B$ & $K'$ & $\alpha_1$ & $\delta$ & $R$ &   Coefficient & Order &  
$\gamma_{br}$ & $B$ & $K'$ & $\alpha_1$ & $\delta$ & $R$ &   Coefficient \\
\hline \hline  \\[-1.0em]
 0 & -   & -  & - & - &  - & - & 6.81783   & 0 & - & - & - & - & - & - & -70.5346 \\
 1 & 1  & -  & - & - &  - & - & 1.99229   & 1 & 1 & - & - & - & - & - & 2.99479\\
 1 & -   & 1 & - & - &  - & - & 0.999460   & 1 & - & 1 & - & - & - & - & 1.99998\\
 1 & -   & -  & - & 1 &  - & - & -0.852880 & 1 & - & - & 1 & - & - & - & 1.00000  \\
 1 & -   & -  & - & - &  1 & - & 1.00159  & 1 & - & - & - & 1 & - & - & 0.220701 \\
 &   &   &    &   &    &   &             & 1 & - & - & - & - & 1 & - & 3.99998\\
 &   &   &    &   &    &   &             & 1 & - & - & - & - & - & 1 & 3.00004\\
\hline
\hline
\end{tabular}
\end{center}
\end{table*}

\begin{table*}[h!]
\begin{center}
\caption[Coefficients of the fit performed to obtain an expression of $\nu F_{\nu;GeV}$ (left) and $\nu F_{\nu;TeV}$ (right) for the case of \tenten.]{Coefficients of the fit performed to obtain an expression of $\nu F_{\nu;GeV}$ (left) and $\nu F_{\nu;TeV}$ (right) for the case of \tenten. The corresponding plots are the central plots of Fig. \ref{1RXSJ1010fits}. For more details see the Caption of Table \ref{cons1010sync}.}
\label{cons1010fgamma}
\begin{tabular} {|c|c|c|c|c|c|c||c|||c|c|c|c|c|c|c||c|}
\hline
\multicolumn{8}{|c}{{$\nu F_{\nu;GeV}$}} & \multicolumn{8}{c|}{{$\nu F_{\nu;TeV}$}}\\
\hline
\hline
Order & $\gamma_{br}$ & $B$ & $K'$ & $\alpha_1$ & $\delta$ & $R$ &   Coefficient & 
Order & $\gamma_{br}$ & $B$ & $K'$ & $\alpha_1$ & $\delta$ & $R$ &   Coefficient \\
\hline \hline  \\[-1.0em]
0 & - & - & - & - &  - & - & -115.648 & 0 & - & - & - & - & - & - & -114.123 \\
 1 & 1 & - & - & - &  - & - & 17.7128   & 1 & 1 & - & - & - & - & - & 13.7218 \\
 2 &2 & - & - & - &  - & - & -1.58660  & 2 & 2 & - & - & - & - & - & -0.850289 \\
 1 &- & 1 & - & - &  - & - & 5.19307   & 1 & - & 1 & - & - & - & - & -4.40501 \\
 2 &1 & 1 & - & - &  - & - & -0.813593   & 2 & 1 & 1 & - & - & - & - & 3.44968  \\
 1 &- & - & 1 & - &  - & - & 1.99993  & 3 & 2 & 1 & - & - & - & - & -0.444764 \\
 1 &- & - & - & 1 &  - & - & 16.9990   & 1 & - & - & 1 & - & - & - & 1.99314\\
 2 &1 & - & - & 1 &  - & - & -14.1166  & 1 & - & - & - & 1 & - & - & -10.8581\\
 3 &2 & - & - & 1 &  - & - & 2.34452   & 2 & 1 & - & - & 1 & - & - & 2.34036\\
 2 &- & 1 & - & 1 &  - & - & -4.22521  & 3 & - & - & - & 3 & - & - & 2.69285\\
 3 &1 & 1 & - & 1 &  - & - & 1.19513   & 3 & - & - & - & - & 1 & - & -13.1497\\
 1 &- & - & - & - &  1 & - & 5.74872   & 2 & 1 & - & - & - & 1 & - & 8.23156\\
 2 &1 & - & - & - &  1 & - & -0.506995 & 3 & 2 & - & - & - & 1 & - & -0.942506\\
 2 &- & - & - & 1 &  1 & - & 1.59946  & 1 & - & - & - & - & - & 1 & 3.98568\\
 1 &- & - & - & - &  - & 1 & 3.99993  &  &  &  &  &  &  &  & \\
\hline
\hline
\end{tabular}
\end{center}
\end{table*}

\begin{table*}[h!]
\begin{center}
\caption[Coefficients of the fit performed to obtain an expression of $\Gamma_{GeV}$ for the case of \tenten.]{Coefficients of the fit performed to obtain an expression of $\Gamma_{GeV}$ for the case of \tenten. The corresponding plot is the left bottom plot of Fig. \ref{1RXSJ1010fits}. For more details see the Caption of Table \ref{cons1010sync}.}
\label{cons1010gFermi}
\begin{tabular} {|c|c|c|c|c|c|c||c|||c|c|c|c|c|c|c||c|}
\hline
\multicolumn{16}{|c|}{{$\Gamma_{GeV}$}}\\
\hline
\hline
Order & $\gamma_{br}$ & $B$ & $K'$ & $\alpha_1$ & $\delta$ & $R$ &   Coefficient & 
Order & $\gamma_{br}$ & $B$ & $K'$ & $\alpha_1$ & $\delta$ & $R$ &   Coefficient \\
\hline \hline  \\[-1.0em] 
0 &  - & - & - & - &  - & - & 215.795    & 2 & 1 & - & - & 1 &  - & - & -16.2468 \\
1 & 1 & - & - & - &  - & - & -202.197    & 3 & 2 & - & - & 1 &  - & - & 1.54404 \\
2 & 2 & - & - & - &  - & - & 67.9303     & 2 & - & 1 & - & 1 &  - & - & -0.383771 \\
3 & 3 & - & - & - &  - & - & -9.87975    & 2& - & - & - & 2 &  - & - & -2.36309  \\
4 & 4 & - & - & - &  - & - & 0.528434    & 1& - & - & - & - &  1 & - & -59.7992\\
1 & - & 1 & - & - &  - & - & -53.6426    & 2& 1 & - & - & - &  1 & - & 39.7326 \\
2 & 1 & 1 & - & - &  - & - & 35.3010     & 3& 2 & - & - & - &  1 & - & -8.64464\\
3 & 2 & 1 & - & - &  - & - & -7.62267    & 4& 3 & - & - & - &  1 & - & 0.621649\\
4 & 3 & 1 & - & - &  - & - & 0.542311    & 3& 1 & 1 & - & - &  1 & - & -0.0933089\\
3 & 1 & 2 & - & - &  - & - & -0.0421037   & 4& 2 & 1 & - & - &  1 & - & 0.017273\\
4 & 2 & 2 & - & - &  - & - & 0.00702904  & 2& - & - & - & 1 &  1 & - & -0.293701\\
1 & - & - & - & 1 &  - & - & 41.7990     & 2& - & - & - & - &  2 & - & -0.0945532\\ 
\hline
\hline
\end{tabular}
\end{center}
\end{table*}

\begin{table*}[h!]
\begin{center}
\caption{Coefficients of the fit performed to obtain an expression of $\Gamma_{TeV}$ for the case of \tenten. The corresponding plot is the right bottom plot of Fig. \ref{1RXSJ1010fits}. For more details see the Caption of Table \ref{cons1010sync}.}
\label{cons1010gHESS}
\begin{tabular} {|c|c|c|c|c|c|c||c|||c|c|c|c|c|c|c||c|} 
\hline
\multicolumn{16}{|c|}{{$\Gamma_{TeV}$}}\\
\hline \hline
Order & $\gamma_{br}$ & $B$ & $K'$ & $\alpha_1$ & $\delta$ & $R$ &   Coefficient & 
Order & $\gamma_{br}$ & $B$ & $K'$ & $\alpha_1$ & $\delta$ & $R$ &   Coefficient \\
\hline \hline  \\[-1.0em]
%\endfirsthead

%\hline
%\hline
%\multicolumn{16}{|r|}
%{{\bfseries \tablename\ \thetable{} -- continued from previous page}} \\ \hline\hline
%Order & $\gamma_{br}$ & $B$ & $K$ & $\alpha_1$ & $\delta$ & $R$ &   Coefficient & 
%Order & $\gamma_{br}$ & $B$ & $K$ & $\alpha_1$ & $\delta$ & $R$ &   Coefficient \tabularnewline
%\hline \hline
%\endhead
%
%\hline
%\multicolumn{16}{|r|}{{Continued on next page}} \\ \hline
%\endfoot
%
%\hline\hline
%\endlastfoot
0 & - & - & - & - &  - & - & 433.763    & 3& - & 1 & - & 1 &  1 & - & -0.688389 \\
1 & 1 & - & - & - &  - & - & 1660.83     & 2& - & - & - & - &  2 & - & -53.2164 \\
2 & 2 & - & - & - &  - & - & -536.572    & 4& 2 & - & - & - &  2 & - & 5.94909 \\
3 & 3 & - & - & - &  - & - & 17.0195     & 5& 3 & - & - & - &  2 & - & -0.738527  \\
1 & - & 1 & - & - &  - & - & 872.651    & 4& 1 & 1 & - & - &  2 & - & 0.943224\\
2 & 1 & 1 & - & - &  - & - & -83.0061     & 5& 2 & 1 & - & - &  2 & - & -0.195947 \\
3 & 2 & 1 & - & - &  - & - & -83.7787     & 3& - & - & - & - &  3 & - & 7.96991\\
4 & 3 & 1 & - & - &  - & - & 5.13618     & 4& 1 & - & - & - &  3 & - & -1.73813\\
5 & 4 & 1 & - & - &  - & - &0.120479    & 1& - & - & - & - &  - & 1 & -254.494\\
2 & - & 2 & - & - &  - & - & -209.046    & 2& 1 & - & - & - &  - & 1 & -102.434\\
3 & 1 & 2 & - & - &  - & - & 112.983    & 3& 2 & - & - & - &  - & 1 & 44.4274\\
4 & 2 & 2 & - & - &  - & - & -18.0429      & 2& - & 1 & - & - &  - & 1 & -213.596\\
5 & 3 & 2 & - & - &  - & - & 1.22248   & 3& 1 & 1 & - & - &  - & 1 & 63.8761\\
1 & - & - & 1 & - &  - & - & 937.346     & 4& 2 & 1 & - & - &  - & 1 & 1.03778\\
2 & 1 & - & 1 & - &  - & - & -215.023    & 3& - & 2 & - & - &  - & 1 & 4.71668\\
3 & 2 & - & 1 & - &  - & - & -17.4660    & 4& 1 & 2 & - & - &  - & 1 & -1.68036\\
2 & - & 1 & 1 & - &  - & - & 50.7081     & 2& - & - & 1 & - &  - & 1 & -122.306\\
3 & 1 & 1 & 1 & - &  - & - & -64.8202    & 3& 1 & - & 1 & - &  - & 1 & 21.0306\\
4 & 2 & 1 & 1 & - &  - & - & 2.94899     & 4& 2 & - & 1 & - &  - & 1 & 3.83924\\
3 & - & 2 & 1 & - &  - & - & 4.10333     & 3& - & 1 & 1 & - &  - & 1 & -13.7376\\
4 & 1 & 2 & 1 & - &  - & - & 2.54005     & 4& 1 & 1 & 1 & - &  - & 1 & 9.62879\\
2 & - & - & 2 & - &  - & - & 43.3093     & 5& 2 & 1 & 1 & - &  - & 1 & -0.221533\\
3 & 1 & - & 2 & - &  - & - & -21.8910    & 4& - & 2 & 1 & - &  - & 1 & -0.296910\\
4 & 2 & - & 2 & - &  - & - & 1.38984    & 5& 1 & 2 & 1 & - &  - & 1 & -0.197426\\
3 & - & 1 & 2 & - &  - & - & -25.7945    & 3& - & - & 2 & - &  - & 1 & -7.67172\\
4 & 1 & 1 & 2 & - &  - & - & 2.88766     & 4& 1 & - & 2 & - &  - & 1 & 2.90132\\
5 & 2 & 1 & 2 & - &  - & - & -0.0347576  & 5& 2 & - & 2 & - &  - & 1 & -0.0925614\\
4 & - & 2 & 2 & - &  - & - & 0.954052    & 4& - & 1 & 2 & - &  - & 1 & 2.60294\\
5 & 1 & 2 & 2 & - &  - & - & -0.0382109  & 5& 1 & 1 & 2 & - &  - & 1 & -0.170639\\
3 & - & - & 3 & - &  - & - & -2.10358   & 5& - & 2 & 2 & - &  - & 1 & -0.0489805\\
4 & 1 & - & 3 & - &  - & - & 0.471742    & 4& - & - & 3 & - &  - & 1 & 0.133559\\
1 & - & - & - & 1 &  - & - & -168.566     & 5& 1 & - & 3 & - &  - & 1 & -0.0299422\\
2 & 1 & - & - & 1 &  - & - & 89.7793    & 2& - & - & - & - &  1 & 1 & -2.83619\\
3 & 2 & - & - & 1 &  - & - & -14.9943     & 3& 1 & - & - & - &  1 & 1 & 0.931345 \\
4 & 3 & - & - & 1 &  - & - & 0.783183    & 3& - & 1 & - & - &  1 & 1 & 0.692465 \\
2 & - & 1 & - & 1 &  - & - & 4.06062    & 3& - & - & 1 & - &  1 & 1 & -0.341208 \\
3 & 1 & 1 & - & 1 &  - & - & -0.0229217     & 4& 1 & - & 1 & - &  1 & 1 & 0.112470  \\
4 & 2 & 1 & - & 1 &  - & - & -0.132376   & 4& - & 1 & 1 & - &  1 & 1 & 0.0839418\\
1 & - & - & - & - &  1 & - & 449.564    & 2& - & - & - & - &  - & 2 & -6.46489 \\
2 & 1 & - & - & - &  1 & - & -44.9238     & 3& 1 & - & - & - &  - & 2 & 4.72897\\
3 & 2 & - & - & - &  1 & - & -76.3805    & 4& 2 & - & - & - &  - & 2 & -1.66802\\
4 & 3 & - & - & - &  1 & - & 21.1307     & 3& - & 1 & - & - &  - & 2 & 7.24229\\
5 & 4 & - & - & - &  1 & - & -1.57986    & 4& 1 & 1 & - & - &  - & 2 & -1.92839\\
2 & - & 1 & - & - &  1 & - & 251.194     & 5& 2 & 1 & - & - &  - & 2 & -0.0907203\\
3 & 1 & 1 & - & - &  1 & - & -182.160    & 3& - & - & 1 & - &  - & 2 & 3.92390\\
4 & 2 & 1 & - & - &  1 & - & 41.0579     & 4& 1 & - & 1 & - &  - & 2 & -0.434856\\
5 & 3 & 1 & - & - &  1 & - & -3.01225    & 5& 2 & - & 1 & - &  - & 2 & -0.177589\\
5 & 2 & 2 & - & - &  1 & - & 0.000509436 & 4& - & 1 & 1 & - &  - & 2 & 0.657782\\
2 & - & - & 1 & - &  1 & - & 4.95581    & 5& 1 & 1 & 1 & - &  - & 2 & -0.337476\\
3 & 1 & - & 1 & - &  1 & - & -1.68709   & 5& - & 2 & 1 & - &  - & 2 & 0.0133100\\
3 & - & 1 & 1 & - &  1 & - & -1.55750   & 4& - & - & 2 & - &  - & 2 & 0.308334\\
4 & 1 & 1 & 1 & - &  1 & - & 0.0483315   & 5& 1 & - & 2 & - &  - & 2 & -0.0937625\\
2 & - & - & - & 1 &  1 & - & 10.2464    & 5& - & 1 & 2 & - &  - & 2 & -0.0619270\\
3 & 1 & - & - & 1 &  1 & - & -2.65305    & &  &  &  &  &   &  & \\
\hline
\hline
\end{tabular}
\end{center}
\end{table*}

  \end{appendix}
    
  \end{document}